# FIELDS, PARTICLES,
## AND NUCLEI

# New Narrow N(1685) and N(1726)? Remarks on the Interpretation of the Neutron Anomaly as an Interference Phenomenon[1]

**V. Kuznetsov**[a, b, *], **V. Bellini**[b, c], **V. Brio**[b, c], **A. Gridnev**[a], **N. Kozlenko**[a], **F. Mammoliti**[b, c],
**F. Tortorici**[b, c], **M. V. Polyakov**[a, d], **G. Russo**[b, c], **M. L. Sperduto**[b, c],
**V. Sumachev**[a], and **C. M. Sutera**[b]

[a] Petersburg Nuclear Physics Institute, Gatchina, 188300 Russia
[b] NFN—Sezione di Catania, Catania, I-95123 Italy
[c] Dipartimento di Fisica ed Astronomia, Università di Catania, Catania, I-95123 Italy
[d] Institute für Theoretische Physik II, Ruhr-Universität Bochum, Bochum, D-44780 Germany
*e-mail: kuznetsov_va@pnpi.nrcki.ru
Received March 22, 2017

Different interpretations of narrow structures at $W \sim 1.68$ and 1.72 GeV observed in several reactions are discussed. It is questionable whether interference phenomena could explain the whole complex of experimental findings. More probable hypotheses would be the existence of one or two narrow resonances $N(1685)$ and $N(1726)$ and/or the sub-threshold virtual $K\Sigma$ and $\omega p$ production (cusps).

The observation of a narrow enhancement at $W \sim 1.68$ GeV in the $\gamma n \rightarrow \eta n$ excitation function [1–7] (the so-called "neutron anomaly") is one of the challenging findings in the domain of hadronic physics. It may signal a new nucleon resonance $N(1685)$ with unusual properties: the narrow ($\Gamma \leq 30$ MeV), strong photo-excitation on the neutron and the suppressed decay to the $\pi N$ final state [8].

On the other hand, several groups explained this enhancement in terms of the interference in the photo-excitation of well-known resonances in $S$ and $P$ waves [9, 10]. For example, the Bonn-Gatchina (BnGa) group suggested that the interference of $S_{11}(1535)$ and $S_{11}(1650)$ may generate this peak [9].

The decisive identification of the neutron anomaly is a challenge for both theory and experiment. Recently the A2@MAMIC Collaboration reported a measurement of the helicity-dependent $\gamma n \rightarrow \eta n$ cross sections $\sigma_{1/2}$ and $\sigma_{3/2}$ [11]. These cross sections correspond to the antiparallel and parallel orientations of the polarizations of the incoming photon and target neutron respectively. The conclusion based on the Legendre-polynomials decomposition was that "… *this structure is related to the helicity-1/2 amplitude and a comparison with different models favours a scenario with a contribution with a narrow $P_{11}$ resonance.*"

This statement was disputed by the authors of [12]. By using the recent solution of the BnGa partial wave analysis, they arrived at the different conclusion: "*There is the suspicion that the dip might be a statistical fluctuation. … A partial wave analysis without a narrow* $J^P = 1/2^+$ resonance is excellent, the inclusion of it with the reported properties leads to a significantly worse description of the ($\gamma n \to \eta n$) data ... our fit without an introduction of a narrow resonance ... returns a $\chi^2 = 1205$ for 1150 data points. Obviously, there is no need to introduce $N(1685)$ ... When $N(1685)$ was enforced ...the fit returned $\chi^2 = 1834$ for 1150 data points..." Definitely, there is a need to examine contradicting assumptions.

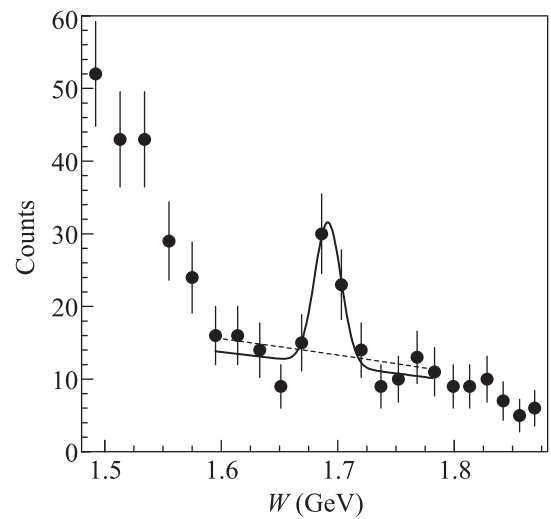

**Fig. 1.** Yield of Compton scattering events on the neutron (figure from [14]). Solid line is the 3d-order polynomial-plus-Gaussian (i.e., background-plus-signal) fit and dashed line indicates 3d-order polynomial fit only. See [14] for details.

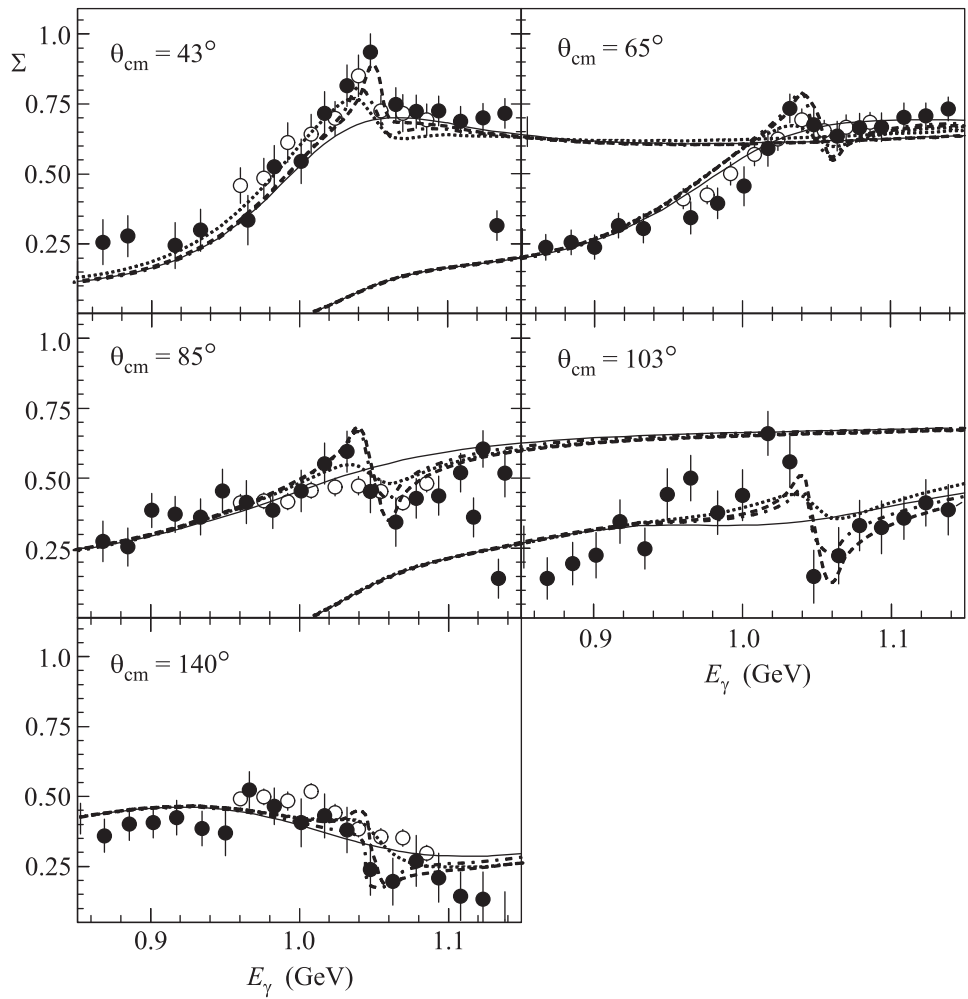

**Fig. 2.** Beam asymmetry for the $\gamma p \to \eta p$ reaction (figure from [15]). Black circles are the data from [15] and open circles are the data from [16]. (Problems in their analysis that led to this conclusion were discussed in detail in [15]. At time of the preparation of this manuscript, this critique remains unreplied.) The solid line is the solution of the SAID partial-wave analysis without a narrow resonance. Dashed, dash-dotted, and dotted lines indicate the same solution with the narrow $P_{11}$, $P_{13}$, and $D_{13}$ resonances, respectively. See [15] for details. It is worth to noting that the authors of this paper arrived at a different conclusion.

First, one should point out one technical problem. In the case of photon interaction with a neutron bound in a deuteron target, Fermi motion of the target neutron changes the effective energy of this interaction and affects momenta of outgoing particles. Some events may suffer from the final-state interaction of reaction products. Kinematics of events is "peaked" around that on a free neutron.

Cuts used for the selection of events in data analyses (e.g., the cut on the neutron missing mass), eliminate not only the background but also those events whose kinematics is stronger distorted by Fermi motion. The quality of the rejection of such events depends on response functions of detectors.

The cuts affect the shape of quasi-free cross section and, in the case of $\gamma n \to \eta n$, may change the width and the position of the observed peak. The tighter cuts make the peak at $W \sim 1.68$ GeV narrower and better pronounced. The influence of the cuts on the $\gamma n \to \eta n$ cross section is discussed in detail in [1].

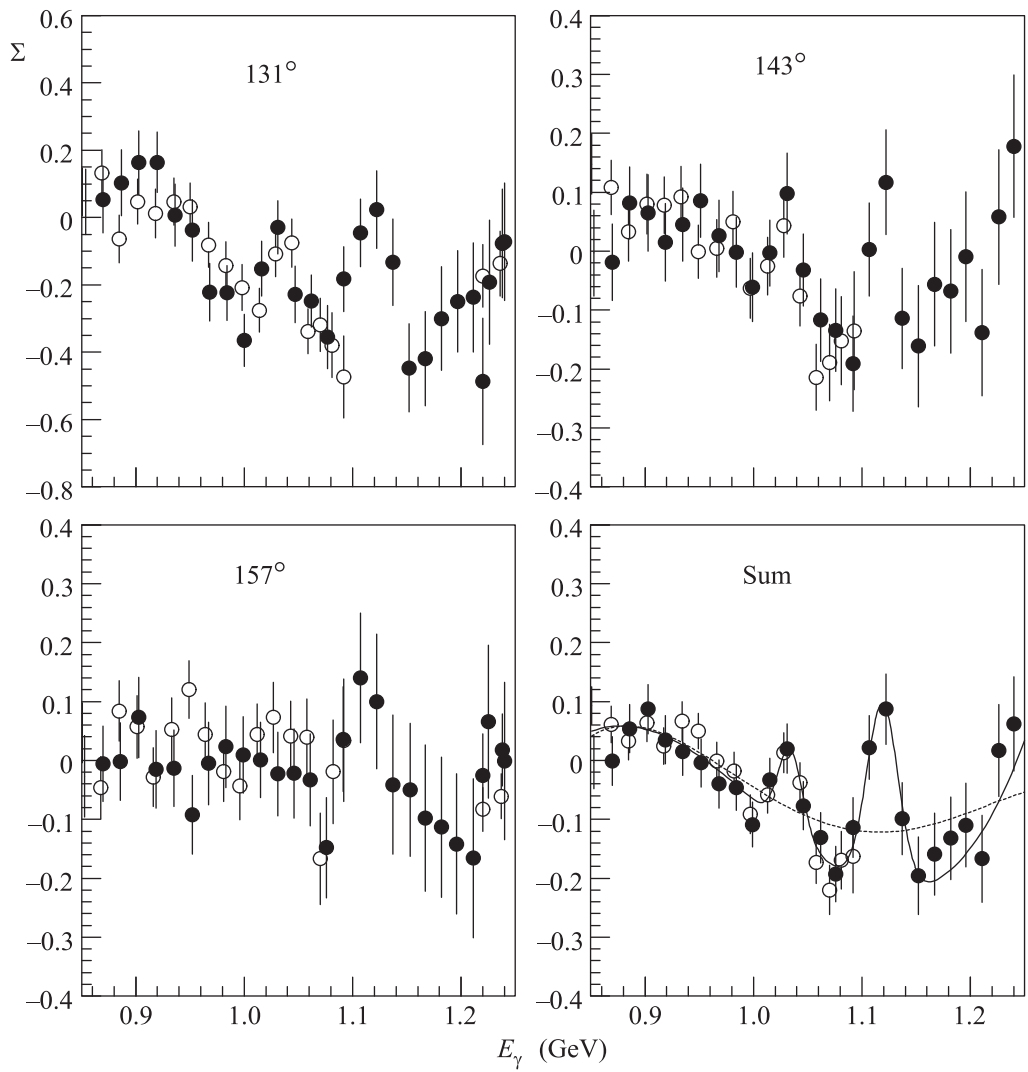

**Fig. 3.** Beam asymmetry for Compton scattering on the neutron (figure from [17]). Solid line is the signal-plus-background fit and dashed line is background fit only. See [17] for details.

That is why the quasi-free $\gamma n \to \eta n$ cross sections reported in [1−7] are unavoidably analysis- and facility-dependent. Could a fit of the $\gamma n \to \eta n$ data, if being done just as cross section calculated for the free neutron and then integrated with the deuteron wavefunction defined by Bonn or Paris potentials (as explained, e.g., in [13]) lead to unambiguous results?

There is a more general consideration. The authors of [9, 10, 12] focused on only the $\gamma n \to \eta n$ cross section, whereas the database for the possible signals of $N(1685)$ is larger. Narrow structures at this energy were also observed in Compton scattering on the neutron $\gamma n \to \gamma n$ [14] (Fig. 1) and in the beam asymmetry for the η photoproduction of the proton $\gamma p \to \eta p$ [15, 16] (Fig. 2). Furthermore, the recent data on the beam asymmetry for Compton scattering on the proton $\gamma p \to \gamma p$ [17], the precise data for the $\gamma n \to \eta n$ [18] and $\pi^- p \to \pi^- p$ [19] reactions revealed two narrow structures at $W \sim 1.68$ and $1.72$ GeV (Figs. 3−5, respectively).

Quote from the paper of the BnGa group [9]: "... *A trace of $N^*(1685)$ may have been found in the $\gamma n \to \gamma n$ total cross section [14] and in the beam asymmetry for $\gamma p \to \gamma p$ [17]. We do not see how the two phenomena could possibly be related to the interference pattern in the $\gamma n \to \eta n$ discussed in this paper*".

Indeed, let us assume that the specific interference of wide ($\Gamma \sim 100{-}200$ MeV) resonances (e.g., $S_{11}(1535)$ and $S_{11}(1650)$) generates a narrow ($\Gamma \le 30$ MeV) peak in the $\gamma n \to \eta n$ excitation function. This reaction is governed by isospin-1/2 resonances only. In contrast, both isospin-1/2 and isospin-3/2 resonances are excited in Compton scattering. The isospin-3/2 resonances (e.g., $D_{33}(1700)$) might be major contributors to this reaction [20].

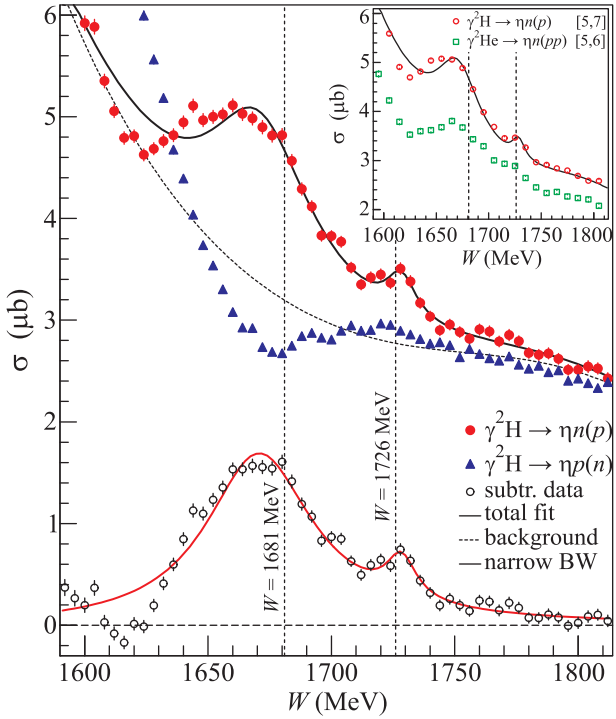

**Fig. 4.** (Color online) Excitation functions for quasi-free photoproduction of η mesons (figure from [18]). Inset: total cross sections for quasi-free neutrons bound in the deuteron (red circles) and in ³He nuclei (green squares). Fit curves include Breit–Wigner resonance for the structure at 1726 MeV. Main figure: finer binned data for the measurement with deuterium data (blue triangles): quasi-free neutron (proton) data. Curves: total fit (solid black curve), background contributions (dashed black curve), and sum of two narrow signal Breit–Wigner functions (red solid curve). Open black circles: background-subtracted neutron data. Vertical lines: markers at $W = 1681$ MeV and $W = 1726$ MeV. See [18] for details.

Could the interference of all these resonances also generate a peak at the same energy in $\gamma n \to \gamma n$? Even assuming that, why this peak is not seen in $\gamma n \to \pi^0 n$ [21], which is governed by the same resonances as Compton scattering?

Could the models [9, 10] explain a structure at $W \sim 1.68$ MeV observed in the beam asymmetry $\Sigma$ for η photoproduction on the proton [15, 16]? The BnGa solution without $N^*(1685)$ is flat [22]. In contrast, the signal of $N(1685)$ may appear in polarization observables for $\gamma p \to \eta p$ even if this resonance only weakly photoexcited on the proton.

Could the interference of wide resonances explain the second structure at $W \sim 1.72$ GeV? Could the BnGa and other solutions reproduce this structure in the $\gamma n \to \eta n$ cross section?

These questions should be answered prior the calculations from [9, 10] (including the BnGa solution used in [12]) could be employed to achieve any meaningful conclusions.

Moreover, in [23] it was shown that the flavor $SU(3)$ symmetry implies that BnGa fine tuning of photocouplings of known $N(1535)$ and $N(1650)$ resonances unavoidably leads to a huge coupling to φ-meson to $N(1650)$ resonance at least 5 times larger than the corresponding $\rho^0$ coupling. In terms of quark degrees of freedom, this means that the well-known $N(1650)$ resonance must be a cryptoexotic pentaquark. Its wavefunction should contain predominantly an $s\bar{s}$ component. It turns out that the conventional interpretation of the neutron anomaly by the interference of known resonances metamorphoses into unconventional physics picture of $N(1650)$.

More natural explanation of two observed phenomena would be the existence of one or two narrow resonances ($N(1685)$ and $N(1726)$). The properties of $N(1685)$, namely the narrow width $\Gamma \leq 30$ MeV, the strong photo-excitation on the neutron and the suppressed decay to the $\pi N$ final state, do coincide well to those expected for the second member of the exotic antidecuplet predicted in the framework of the Chiral Soliton Model [24] (pentaquarks). However, its decisive accusation requires in particular the identification of the second structure at $W \sim 1.726$ GeV.

Another assumption would be the sub-threshold virtual $K\Sigma$ and $\omega p$ productions (cusps). It is favored by the fact that both structures are observed at the energies which correspond to the thresholds of these reactions. One should understand why these cusps are seen in the η photoproduction and Compton scattering on the neutron and are not (or poorly) seen in these reactions on the proton and in the $\pi^0$ photoproduction on both the neutron and the proton. Only one model-dependent calculation of the sub-threshold $K\Sigma$ photoproduction [25] is now available.

In summary, it looks questionable whether the interference phenomena could accommodate all experimental observations. Two other hypotheses, namely the existence of one or two narrow resonances or cusps, require further theoretical and experimental studies.

This work was supported by the High Energy Department, Petersburg Nuclear Physics Institute; by the INFN Section of Catania; and by the Ruhr Uni-versity of Bochum. We are very grateful to Prof. M. Oleynik for the assistance in preparation of this manuscript.

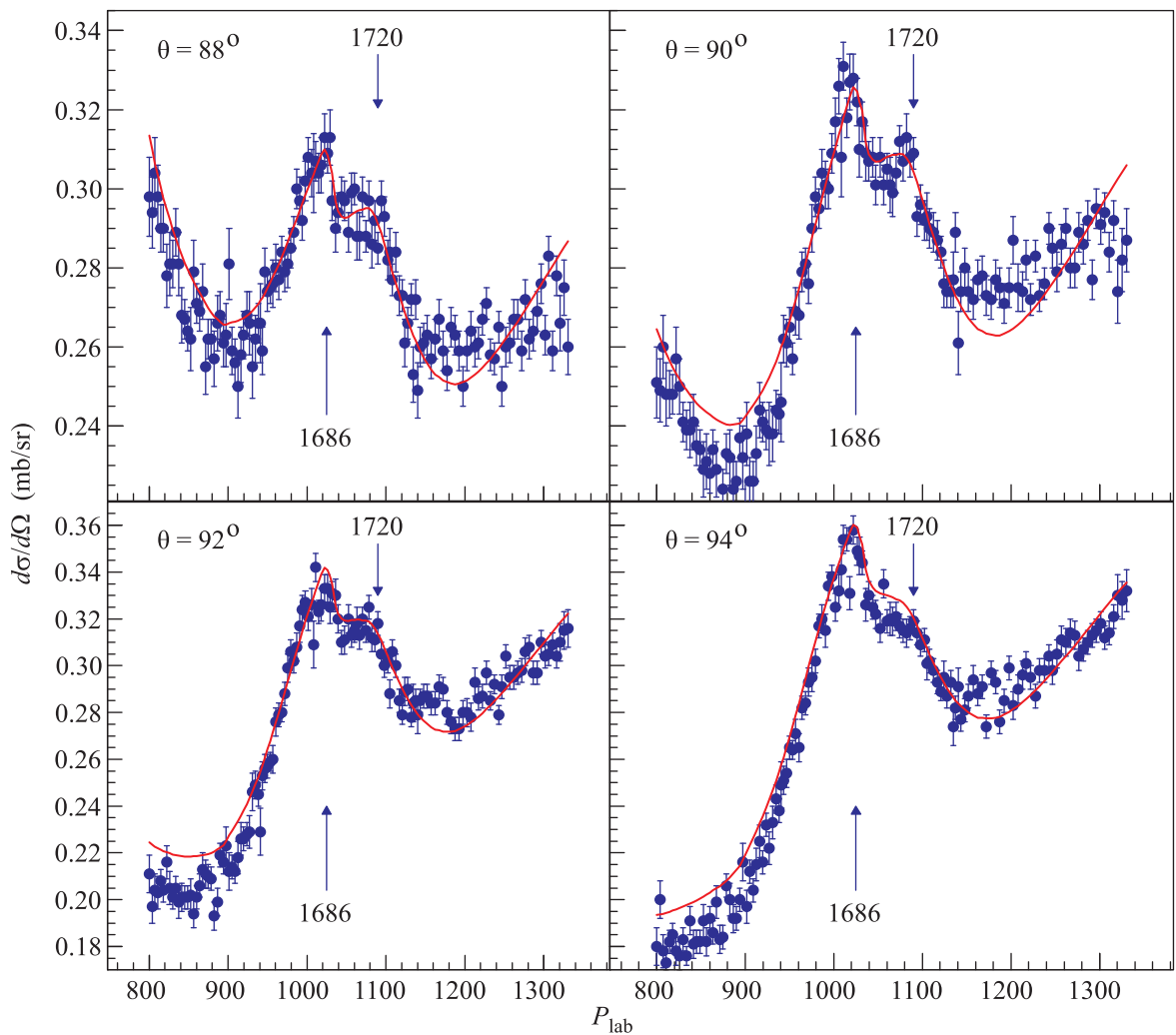

**Fig. 5.** (Color online) $\pi^- p \to \pi^- p$ cross section (figure from [19]). Red solid lines correspond to the calculations based on the $K$-matrix approach with two narrow resonances. See [19] for details.

SPELL: 1. favours